\documentclass[fleqn,usenatbib]{mnras}

\usepackage{newtxtext,newtxmath}

\usepackage[T1]{fontenc}

\DeclareRobustCommand{\VAN}[3]{#2}
\let\VANthebibliography\thebibliography
\def\thebibliography{\DeclareRobustCommand{\VAN}[3]{##3}\VANthebibliography}

\usepackage{graphicx}	
\usepackage{amsmath}	

\title[Turbulent Magnetic Field Line Separation by RBD]{Magnetic Field Line Separation by Random Ballistic Decorrelation in Transverse Magnetic Turbulence}

\author[C. Yannawa et al.]{
C. Yannawa,$^{1}$
P. Pongkitiwanichakul,$^{1}$\thanks{E-mail: peera.po@ku.th}
D. Ruffolo,$^{2}$
P. Chuychai,$^{3}$
and W. Sonsrettee$^{4}$
\\
$^{1}$Department of Physics, Faculty of Science, Kasetsart University, Bangkok 10900, Thailand\\
$^{2}$Department of Physics, Faculty of Science, Mahidol University, Bangkok 10400, Thailand\\
$^{3}$33/5 Moo 16, Tambon Bandu, Muang District, Chiang Rai 57100, Thailand\\
$^{4}$Faculty of Engineering and Technology, Panyapiwat Institute of Management, Nonthaburi 11120, Thailand
}

\pubyear{2023}

\begin{document}
\label{firstpage}
\pagerange{\pageref{firstpage}--\pageref{lastpage}}
\maketitle

\begin{abstract}
The statistics of the magnetic field line separation provide insight into how a bundle of field lines spreads out and the dispersion of non-thermal particles in a turbulent environment, which underlies various astrophysical phenomena. Its diffusive character depends on the distance along the field line, the initial separation, and the characteristics of the magnetic turbulence. This work considers the separation of two magnetic field lines in general transverse turbulence in terms of the magnetic power spectrum in three-dimensional wavenumber space. We apply non-perturbative methods using Corrsin’s hypothesis and assume random ballistic decorrelation to calculate the ensemble average field line separation for general transverse magnetic turbulence. For 2D+slab power spectra, our analytic formulae and computer simulations give similar results, especially at low slab fraction. Our analytical expression also demonstrates several features of field line separation that are verified by computer simulations.
\end{abstract}




\section{Introduction}

Magnetic field lines in turbulent plasmas have trajectories governed by a random-walk process. These field lines determine spatial connectivity for charged components of the plasma and channel nonthermal particles, such as solar energetic particles, toward broader regions as they spread out \citep{Jokipii1966, Tooprakai2016,Laitinen2023}. As the field lines diffuse, charged particles basically follow the field lines \citep{Minnie2009} and thereby undergo diffusion perpendicular to the mean magnetic field in astrophysical and space plasmas \citep[e.g.,][]{Matthaeus2003,Shalchi2006,Ruffolo2012,Shalchi2020,Shalchi2021}, and confined plasmas such as in a tokamak \citep[e.g.,][]{ Rechester&Rosenbluth1978,Kadomtsev&Pogutse1979, Isichenko1991a,Isichenko1991b}.

Many works have developed models to estimate the diffusion coefficient of magnetic field lines in a turbulent environment, as recently reviewed by \cite{Shalchi2021a} and \cite{Engelbrecht2022}.
A perturbative method or a quasilinear theory, similar to a traditional random walk, provides the diffusion coefficient proportional to the square of the magnetic fluctuation \citep{Jokipii&Parker1968}. This theory works well for fluctuations that predominantly vary along the mean magnetic field (with power mainly at wavevectors parallel to the mean field) and are exact for slab fluctuations that vary exclusively along the mean field. There are also non-perturbative calculations (e.g., by \cite{Kadomtsev&Pogutse1979} for an special case) and methods that use Corrsin's hypothesis \citep{Corrsin59} to consider the distribution in transverse displacement of magnetic field lines \citep{Matthaeus1995}.
As confirmed by numerical simulations, such methods work well over wide parameter regimes \citep{Ghilea2011,Snodin2016}.

Using Corrsin's hypothesis requires specifying the probability distribution of the transverse displacement as mentioned above. One popular probability form is Gaussian, and its variance must be specified. Diffusive decorrelation (DD) \citep{Matthaeus1995,Ruffolo2004} and random ballistic decorrelation (RBD) \citep{Ghilea2011} have been proposed as analytic forms for this variance. While a self-consistent form has also been proposed \citep{Saffman1963,TaylorandMcNamara1971,ShalchiandKourakis2007}, this requires solving an ordinary differential equation (ODE), is computationally expensive, and does not necessarily improve the results \citep{Snodin2013}. By using DD, the field line displacement is assumed to be diffusive at all length scales. However, as diffusive behavior is unrealistic at scales smaller than the correlation length, DD may not provide a good approximation over distances smaller than the correlation scale, which provide the dominant contribution to the calculation of diffusion coefficients. 
Another disadvantage of using DD is that the diffusion coefficient is provided in an implicit form, which is less easily solved than an explicit form.
In contrast, the RBD assumption sets field line trajectories to be ballistic, along straight lines. Such an assumption is valid for the small displacements that dominate the calculation of field line displacement statistics.
An advantage of using RBD is that the solution is in an explicit form that can be solved directly. In any case, for the field line random walk, the DD, RBD, and ODE formulations all give asymptotic (long-distance)  field line diffusion coefficients in good agreement with computer simulations for a wide parameter range \citep{Ghilea2011,Snodin2016}.

As field lines wander and diffuse in space, a group of previously close field lines can expand, spread, and mix with others. How large such a group becomes influences the spread of energetic particles, which for example can complicate efforts to determine the sources of such particles.
To understand such expansion, the separation of two nearby field lines \citep{Jokipii1973,Ruffolo2004,Shalchi2019} has been investigated as a function of distance along the mean magnetic field.

In this work, we extend the analysis of \cite{Ruffolo2004} who used the DD approximation to examine the separation of magnetic field lines. 
We consider general transverse magnetic turbulence, and we replace the DD assumption with RBD and provide an expression for the separation of two field lines as a function of the distance along the mean field.
The restriction to transverse magnetic fluctuations, which have been called ``incompressible,'' can be motivated by observations of magnetic fluctuations in the solar wind, in which $\approx90$\% of the fluctuation energy comes from transverse components ~\citep{Belcher&Davis1971}.
This form allows for any specification of magnetic spectra. 
We compare our RBD results with computer simulations and with results from the DD theory of \cite{Ruffolo2004} for 2D+slab magnetic turbulence. We find that our RBD and DD formulae both provide good approximations to diffusion coefficients over an extensive range of conditions. Furthermore, RBD enables an explicit solution for diffusion coefficients that is more straightforward than the implicit solution from the DD approach.

\begin{figure*}
   \centering
   \includegraphics[width=0.85\textwidth]{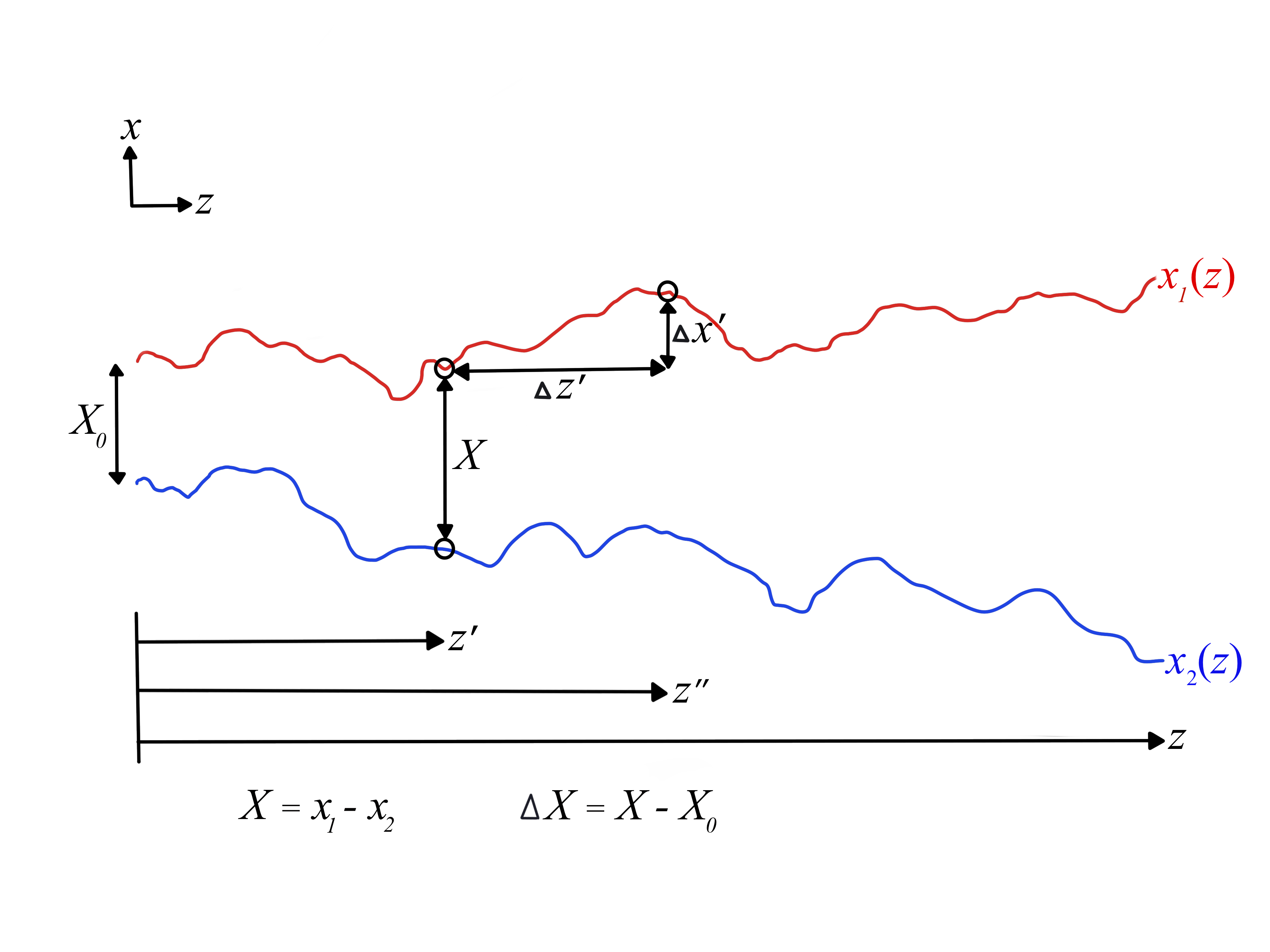}
      \caption{Illustration of separation of two nearby magnetic field lines in the $x$-$z$ plane where $z$-axis is along the mean field and $x$-axis is along the initial displacement, which has magnitude $X_{0}$. After the change in $z$ indicated by $\Delta z^{\prime} = z^{\prime \prime}-z^{\prime}$, the change in $\bf{x}$ of a single field line is $\bf{\Delta x^{\prime}}$. The displacement between the two field lines is indicated by $\bf{X}=\bf{x_1}-\bf{x_2}$ and the separation is $\bf{\Delta X} = \bf{X}-\bf{X_0} $. In the present work we calculate $\langle \Delta X^{2} \rangle$ and $\langle \Delta Y^{2} \rangle$ as functions of $z$.
              }
         \label{fig:fls}
   \end{figure*}

\section{Theory of Field Line Separation for General Transverse Magnetic Turbulence}

We follow the derivation of~\cite{Ruffolo2004} by considering magnetic fluctuations in the $x$-$y$ plane, which are transverse to the mean magnetic field along the $\hat{z}$ direction. The total magnetic field is
\begin{equation}
{\bf B} = b_{x}(x,y,z){\bf\hat{x}}+b_{y}(x,y,z){\bf\hat{y}}+B_{0}{\bf\hat{z}},
\end{equation}
where $B_0$ is the magnitude of the mean magnetic field and $b_x$ and $b_y$ are $x$ and $y$ components of the transverse fluctuating magnetic field. The field line separation depends on the statistics of the field line random walk; therefore, we begin by introducing the Corrsin-based field line random walk theory~\citep{Matthaeus1995,Ruffolo2004,Ghilea2011}.
The set of equations for the field line trajectory is

\begin{equation}
    \frac{dx}{b_{x}(x,y,z)} = \frac{dy}{b_{y}(x,y,z)} = \frac{dz}{B_0}.
    \label{eq:fieldline}
\end{equation}
The displacement $\Delta{x}$ along the $x$ direction after a distance $z$ along the mean field direction is
\begin{equation}
    \Delta x  = \frac{1}{B_0} \int_0^{z}  b_x[x({z}'),y({z}'),z']d{z}'
\end{equation} 
and the ensemble average variance of $\langle\Delta x^2\rangle$ can be written as \citep{Ruffolo2015}
\begin{eqnarray}
    \langle\Delta x^{2}\rangle &=&\frac{1}{B_{0}^2}\int_{0}^{z}\int_{0}^{z}\langle b_{x}(x',y',z')b_{x}(x'',y'',z'')\rangle dz'dz'',\nonumber \\
\end{eqnarray}
where $x'\equiv x(z')$ and $y'\equiv y(z')$. Without loss of generality we can redefine $z'$ and $z''$ such that  $z''\geq z'$ so 
\begin{eqnarray}
\left \langle \Delta x^2\right \rangle &=&\frac{2}{B_0^2} \int_0^z\int_0^{z-{z}'} \left\langle b_x(0,0,0)b_x(\Delta x',\Delta y',\Delta z') \right \rangle_L 
\nonumber \\
& & \times d\Delta z' d{z}'.
\end{eqnarray}
We have assumed that the statistics of the correlation are homogeneous, $\Delta x'\equiv x'' - x'$, $\Delta y'\equiv y'' - y'$, and $\Delta z'\equiv z'' - z'$.
The symbol $\langle ... \rangle$ represents an ensemble average. The subscript $L$ stands for Lagrangian, meaning that quantities must be calculated along the same field line.  
We use Corrsin's hypothesis to write the Lagrangian ensemble average in terms of the Eulerian ensemble average
$R_{xx}(\Delta x',\Delta y',\Delta z') \equiv \left \langle b_x (0,0,0)b_x (\Delta x',\Delta y',\Delta z') \right \rangle$ as
\begin{eqnarray}
\left\langle b_x(0,0,0)b_x(\Delta x',\Delta y',\Delta z') \right \rangle_L &=& \int_{-\infty}^\infty \int_{-\infty}^\infty R_{xx}(\Delta x',\Delta y',\Delta z') \nonumber \\
& & \times P(\Delta x' \mid \Delta z')P(\Delta y' \mid \Delta z') \nonumber \\
& & \times d\Delta x' d\Delta y' ,
\end{eqnarray}
where $P(\Delta x'\mid\Delta z')$ and $P(\Delta y' \mid \Delta z')$ are the probabilities that a field line has displacement components of $\Delta x'$ and $\Delta y'$, respectively, over the distance $\Delta z'$. 
We note that
\begin{eqnarray}
R_{xx}(\Delta x',\Delta y',\Delta z') &=& \frac{1}{(2\pi)^{3/2}}{\int} P_{xx}(\mathbf{k})
e^{-ik_x\Delta x'} e^{-ik_y\Delta y'} \nonumber \\
& & \times e^{-ik_z\Delta z'} d\mathbf{k}
\label{eq:euler}
\end{eqnarray}
in terms of magnetic power spectra.
Using Corrsin's hypothesis gives
\begin{eqnarray}
\left \langle \Delta x^2 \right \rangle &=& \frac{2}{B_0^2} \int_0^z\int_0^{z-{z}'}\int_{-\infty}^\infty \int_{-\infty}^\infty R_{xx}(\Delta x',\Delta y',\Delta z') \nonumber \\
& & \times P(\Delta x' \mid \Delta z') P(\Delta y' \mid \Delta z') d\Delta x' d\Delta y'  d\Delta z' d{z}'.
\label{eq:x2}
\end{eqnarray}
$P(\Delta x' \mid \Delta z')$ and $P(\Delta y' \mid \Delta z')$ are assumed to be Gaussian,
\begin{eqnarray}
     P(\Delta x' \mid \Delta z') &=& \frac{1}{\sqrt{2 \pi \sigma_x^2}} \exp \left [ \frac{-(\Delta x')^2}{2\sigma_x^2} \right ],\nonumber \\
     P(\Delta y' \mid \Delta z') &=&\frac{1}{\sqrt{2 \pi \sigma_y^2}} \exp \left [ \frac{-(\Delta y')^2}{2\sigma_y^2} \right ],
\end{eqnarray}
where $\sigma_x^{2}(\Delta z')$ and $\sigma_y^{2}(\Delta z')$ are the variances. By using RBD, we assume that $\Delta x'=(b_{x}/B_{0})\Delta z'$ and $\Delta y'=(b_{y}/B_{0})\Delta z'$. 
The variances are then set to
\begin{equation}
    \sigma^2_x = \frac{\left \langle b_x^2 \right \rangle}{B_0^2} \Delta z'^{2} 
    \label{eq:Px}
\end{equation}   
and 
\begin{equation}
    \sigma^2_y = \frac{\left \langle b_y^2 \right \rangle}{B_0^2} \Delta z'^{2}
    \label{eq:Py}
\end{equation}
By substituting $R_{xx}(\Delta x',\Delta y',\Delta z')$, $P(\Delta x' \mid \Delta z')$ and $P(\Delta y' \mid \Delta z')$ from Equations~\ref{eq:euler},~\ref{eq:Px}, and~\ref{eq:Py}, respectively, into Equation~\ref{eq:x2}, and integrating over $\Delta x'$ and $\Delta y'$, we get
\begin{eqnarray}
    \left \langle \Delta x^2 \right \rangle &=& \frac{1}{(2\pi)^{3/2}}\frac{2}{B_0^2} {\int}\int_0^z\int_0^{z-{z}'}
    P_{xx}(\mathbf{k}) e^{-ik_z\Delta z'} \nonumber \\
    & & \times e^{-(\langle b^2_x\rangle k_x^2 +\langle b^2_y\rangle k_y^2)\Delta z'^{2}/(2B_{0}^{2})} d\Delta z' d{z}' d\mathbf{k}.
    \label{eq:x2_2}
\end{eqnarray}
We can obtain an analogous expression for $\langle\Delta y^{2}\rangle$ by changing $P_{xx}$ to $P_{yy}$. Next, these expressions are used for the field line separation. We can define the FLRW diffusion coefficient as $D_{x}=\langle \Delta x^{2}\rangle/(2z)$, which tends to a constant at large $z$, indicating a diffusive random walk.

We now consider the separation between two field lines as a function of $z$ and the initial displacement ${\bf X_0}$, which without loss of generality we take to be along the $\hat x$ direction, so that ${\bf X_{0}}=X_{0}\hat x$ (Figure~\ref{fig:fls}). The field line separation in the $x$-direction is defined by $\Delta X \equiv \Delta x_1-\Delta x_2 = X-X_{0}$ and
we can write:
\begin{eqnarray}
    \left \langle \Delta X^2 \right \rangle &=& 2I_{x}-2J_{x},
    \label{eq:DX2_1st}
\end{eqnarray}
where 
\begin{eqnarray}
I_{x} &=& \left \langle \Delta x^2 \right \rangle \\
J_{x} &=& \left \langle \Delta x_1 \Delta x_2 \right \rangle \nonumber \\
    &=& \frac{1}{B_0^2}\left[ \int_0^z \int_{z'}^z \left \langle b_x(x'_1, y'_1, z')b_x(x''_2, y''_2, z'') \right \rangle_L d{z}''d{z}' \right. \nonumber \\
    & & \left. + \int_0^z \int_{z'}^z \left \langle b_x(x'_2, y'_2, z')b_x(x''_1, y''_1, z'') \right \rangle_L d{z}''d{z}'
    \right], \nonumber \\
    \label{eq:X2}
\end{eqnarray}
where we again require $z'\leq z''$. 
This requirement, taking into account the concerns in Footnote 4 of \citet{Ruffolo2015}, represents a slight deviation from the derivation of \citet{Ruffolo2004}. To achieve this, when $z'>z''$, we switch the labels of $z'$ and $z''$, resulting in the second integral. The subscripts $1$ and $2$ are the labels of the field lines 1 and 2. Here we use $\langle \Delta x^2 \rangle=\langle \Delta x_2^2 \rangle = \langle \Delta x_1^2 \rangle$, and note that $I_{x}=2D_{x}z$. 
Let us set 
$X \equiv x'_1 - x'_2$, and
$Y \equiv y'_1 - y'_2$ to rewrite  equation~\ref{eq:X2} as
\begin{eqnarray}
    J_{x} &=& \frac{1}{B_0^2} \int_0^z \int_0^{z-z'} \left[ \left \langle b_x(0,0,0)b_x(\Delta x' - X, \Delta y' - Y, \Delta z') \right \rangle_L \right. \nonumber \\
    & & \left. + \left \langle b_x(0,0,0)b_x(\Delta x' + X, \Delta y' + Y, \Delta z') \right \rangle_L \right ] d\Delta z' d{z}'. 
     \label{eq:X2_2}
\end{eqnarray}
Note that homogeneity is assumed to obtain the above equation.
To write $\langle b_x(0,0,0)b_x(\Delta x' \pm X, \Delta y' \pm Y, \Delta z') \rangle_L$ in terms of the Eulerian form $R_{xx}(\Delta x \pm X, \Delta y \pm Y, \Delta z)$, noting that $X=X_{0}+\Delta X $ and $Y=\Delta Y$, we again use Corrsin's hypothesis, which now requires two additional probabilities $P(\Delta X|z')$ and $P(\Delta Y|z')$  of finding the separation of two field lines by $\Delta X$ and $\Delta Y$ over the distance $z'$. These probabilities are assumed to be in the Gaussian forms

\begin{eqnarray}
     P(\Delta X \mid z') &=& \frac{1}{\sqrt{2 \pi \sigma_X^2}} \exp \left [ \frac{-(\Delta X)^2}{2\sigma_X^2} \right ],\nonumber \\
     P(\Delta Y \mid z') &=&\frac{1}{\sqrt{2 \pi \sigma_Y^2}} \exp \left [ \frac{-(\Delta Y)^2}{2\sigma_Y^2} \right ],
\label{eq:PP}
\end{eqnarray}
where
\begin{eqnarray}
    \sigma_X^2 &=&\langle \Delta X^2 \rangle =  \mathcal{D}_x z^{\prime2},
\label{eq:sigma_x} \\
    \sigma_Y^2 &=& \langle \Delta Y^2 \rangle = \mathcal{D}_y z^{\prime2}.
\label{eq:sigma_y} 
\end{eqnarray}
These forms of $\sigma_X$ and $\sigma_Y$ are from RBD. The dimensionless constants $\mathcal{D}_x$ and $\mathcal{D}_y$ can be evaluated by solving for $\langle \Delta X^2 \rangle$ and $\langle \Delta Y^2 \rangle$ in the limit $z \to 0$. Their calculations are shown later.
In terms of the power spectrum, $\langle b_x(0,0,0)b_x(\Delta x' \pm X, \Delta y' \pm Y, \Delta z') \rangle_L$ becomes
\begin{eqnarray}
& & \langle b_x(0,0,0)b_x(\Delta x'\pm X, \Delta y' \pm Y, \Delta z') \rangle_L \nonumber \\
& & = \frac{1}{(2\pi)^{3/2}}{\int} \int_{-\infty}^\infty\int_{-\infty}^\infty \int_{-\infty}^\infty \int_{-\infty}^\infty P_{xx}(\mathbf{k})e^{-ik_x(\Delta x'\pm X)} \nonumber \\
& & \times e^{-ik_y(\Delta y'\pm Y)} e^{-ik_z\Delta z'}  P(\Delta x' \mid \Delta z')P(\Delta y' \mid \Delta z') P(\Delta X\mid z')
\nonumber \\
& & \times P(\Delta Y\mid z') d\Delta x' d\Delta y' dX dY d\mathbf{k}.
\label{eq:Rxx2}
\end{eqnarray}
We substitute $\langle b_x(0,0,0)b_x(\Delta x' - X, \Delta y' - Y, \Delta z') \rangle_L$ and $\langle b_x(0,0,0)b_x(\Delta x' + X, \Delta y' + Y, \Delta z') \rangle_L$ from Equation~\ref{eq:Rxx2} into Equation~\ref{eq:X2_2} and integrate over $\Delta x'$, $\Delta y'$, $X$ and $Y$ to get
\begin{eqnarray}
J_{x} &=& \frac{2}{B_0^2(2\pi)^{3/2}}{\int} \int_0^z \int_0^{z-z'} P_{xx}(\mathbf{k})e^{-ik_z \Delta z'} \nonumber \\
&& \times e^{-(\langle b^2_x\rangle k_x^2 + \langle b^2_y\rangle k_y^2) \Delta z'^{2}/(2B_{0}^{2})} \cos \left (k_x X_0 \right) \nonumber \\
&& \times e^{-(\mathcal{D}_x k_x^2 + \mathcal{D}_y k_y^2) z'^{2}/2} d\Delta z' dz' d\mathbf{k}.
\label{eq:X2_3}
\end{eqnarray}
Substituting $ \langle \Delta x^2 \rangle$ from Equation~\ref{eq:x2_2} and $J_{x}$ from Equation~\ref{eq:X2_3} into Equation~\ref{eq:DX2_1st} gives the expression for $ \langle\Delta X^2 \rangle$:
\begin{eqnarray}
\left\langle \Delta X^2 \right \rangle &=& 2I_{x}-2J_{x} \nonumber \\
&=&
\frac{4}{B_0^2(2\pi)^{3/2}}{\int} \int_0^z \int_0^{z-z'} P_{xx}(\mathbf{k})e^{-ik_z \Delta z'} \nonumber \\
&& \times e^{-(\langle b^2_x\rangle k_x^2 + \langle b^2_y\rangle k_y^2) \Delta z'^{2}/(2B_0^2)} \nonumber \\
&& \times \left[1- \cos \left (k_x X_0 \right) e^{-(\mathcal{D}_x k_x^2 + \mathcal{D}_y k_y^2) z'^2/2}\right ] d\Delta z' dz' d\mathbf{k}.\nonumber \\
\label{eq:X2_4}
\end{eqnarray}
We can obtain an analogous expression for $\Delta Y^2$ by changing $P_{xx}$ to $P_{yy}$. 

To solve for $\mathcal{D}_x$ and $\mathcal{D}_y$, we take the limit of $\left\langle \Delta X^2 \right \rangle$ and $\left\langle \Delta Y^2 \right \rangle$ as $z$ approaches zero and obtain
\begin{equation} 
\mathcal{D}_{x}=\frac{\langle \Delta X^{2} \rangle}{z^2} = 
\frac{2}{(2\pi)^{3/2} B^2_0}{\int}P_{xx}(\mathbf{k})\left [1- \cos \left (k_x X_0\right) \right ]d\mathbf{k}
\label{eq:dX2_zsmall} \\
\end{equation}
and
\begin{equation}
\mathcal{D}_{y}=\frac{\langle \Delta Y^{2} \rangle}{z^2} = 
\frac{2}{(2\pi)^{3/2} B^2_0}{\int}P_{yy}(\mathbf{k})\left [1- \cos \left (k_x X_0 \right) \right ]d\mathbf{k}.
\label{eq:dY2_zsmall} \\
\end{equation}
An alternative way to get $\mathcal{D}_x$ is by considering the separation of two field lines at $z\ll l_\perp$, where $l_\perp$ is a perpendicular bendover scale, for which $\Delta x \approx (b_{x}/B_{0})z$.  Using the initial separation of $X_0$ along the $x$ direction and the definition of $\langle \Delta X^2 \rangle$, we write
\begin{eqnarray}
\mathcal D_x &=&\lim_{z\to 0} \frac{\langle(\Delta x_2-\Delta x_1)^2\rangle}{z^2} \nonumber \\
&=&\frac{1}{B_0^2}\left\langle[b_x(0,0,0)-b_x(X_0,0,0)]^2\right\rangle
\label{eq:actual_X2}
\end{eqnarray}
and similarly
\begin{eqnarray}
\mathcal D_y =\frac{1}{B_0^2}\left\langle[b_y(0,0,0)-b_y(X_0,0,0)]^2\right\rangle.
\label{eq:actual_Y2}
\end{eqnarray}
These are normalized second-order structure functions of the magnetic fluctuations and are exactly equivalent to $\mathcal D_x$ and $\mathcal D_y$ from Equations~\ref{eq:dX2_zsmall} \& \ref{eq:dY2_zsmall} when homogeneity is assumed.  Therefore, $\mathcal D_x$ and $\mathcal D_y$ are small when $X_0\ll l_\perp$, rising toward constant values for larger $X_0$. Note that for $X_{0}\lesssim l_{\perp}$, $\mathcal{D}_x$ and $\mathcal{D}_y$ are generally quite different, as will be shown later. For the special case of axisymmetric 2D+slab spectra, $\mathcal D_x = \mathcal D_y \approx(b^{2D}/B_0)^2$ for $X_0 \gg l_\perp$ because the slab field remains unchanged for the transverse displacement. To solve for $ \langle\Delta X^2 \rangle$ at any $z$, we solve Equation~\ref{eq:X2_4} by using $\mathcal{D}_x$ and $\mathcal{D}_y$ from Equations~\ref{eq:dX2_zsmall} \& \ref{eq:dY2_zsmall}. For solving $\langle\Delta Y^2 \rangle$ we follow the same procedure as for $ \langle\Delta X^2 \rangle$ but replace $P_{xx}$ with $P_{yy}$. 

\section{2D + slab Turbulence}
In this section, we apply our field-line separation formula for 2D+slab axisymmetric turbulent spectra \citep{Bieber1996} and compare our calculation with the field-line separation from the computer simulations and also a DD formulation. According to \cite{Ruffolo2004}, the magnetic fluctuation is $\mathbf{b} = \mathbf{b}^{2D}(x,y) + \mathbf{b}^{slab}(z)$, where $\mathbf{b^{2D}}=\mathbf{\nabla}\times[a(x,y)\hat z]$ to preserve ${\bf \nabla \cdot B}=0$, and  $a(x,y)$ denotes a scalar magnetic potential. Both components are transverse to $B_0\hat{z}$. The general forms of transverse power spectra of the 2D+slab magnetic field are 

\begin{equation}
    P_{xx}(\mathbf{k}) = 2\pi P_{xx}^{slab}(k_z) \delta (k_x) \delta (k_y)
    +\sqrt{2\pi}P_{xx}^{2D}(k_x,k_y) \delta (k_z)
    \label{eq:Px2d}
\end{equation}
and
\begin{equation}
    P_{yy}(\mathbf{k}) = 2\pi P_{yy}^{slab}(k_z) \delta (k_x) \delta (k_y)
    +\sqrt{2\pi}P_{yy}^{2D}(k_x,k_y) \delta (k_z),
    \label{eq:Py2d}
\end{equation}
where $P_{xx}^{2D}$ and $P_{yy}^{2D}$ are related by
\begin{equation}
    P_{xx}^{2D}(k_x,k_y) = k_y^2A(k_x,k_y)
    \label{eq:spec1}
\end{equation}
and
\begin{equation}
    P_{yy}^{2D}(k_x,k_y) = k_x^2A(k_x,k_y),
     \label{eq:spec2}
\end{equation}
where $A(k_{x},k_{y})$ is the Fourier transform of the magnetic potential $a(x,y)$.

Applying these spectra in Equation~\ref{eq:X2_4} and integrating over $\Delta z'$, we get
\begin{eqnarray}
\left\langle \Delta X^2 \right \rangle &=& \frac{2}{\sqrt{\pi} bB_0}\int_{-\infty}^{\infty}\int_{-\infty}^{\infty} \int_0^z \frac{P_{xx}^{2D}(k_x,k_y)}{k_\perp} \nonumber \\
&& \times {\rm erf} \left[\frac{1}{2}\frac{b}{B_0}k_\perp(z-z')\right]\nonumber \\ & & \times \left[1- \cos\left(k_x X_0 \right)e^{-(\mathcal{D}_x k_x^2 + \mathcal{D}_y k_y^2) z'^2/2}\right] dz' dk_x dk_y\nonumber \\
\label{eq:X2_2d+slab_1}
\end{eqnarray}
where $k_{\perp}=\sqrt{k_{x}^{2}+k_{y}^{2}}$. 
Here we have applied the assumption of statistical axisymmetry to set $\langle b_{x}^2 \rangle = \langle b_{y}^2 \rangle = b^{2}/2$, where we define $b^{2}\equiv \langle b^{2} \rangle$.

Note that the slab component of turbulence does not directly contribute to $\langle\Delta X^{2}\rangle$. Physically, this is because slab fluctuations depend only on $z$ and therefore do not cause magnetic field lines to separate, i.e., at a given $z$ the two field lines feel the same slab fluctuation, which makes them move together without contributing to the separation. Mathematically, the slab power is non-zero only at $k_{x}=k_{y}=0$, where the bracketed terms in Equation~\ref{eq:X2_4} sum to zero, i.e., $I_{x}$ and $J_{x}$ cancel out.

As $z\rightarrow 0$, the limit of Equation~\ref{eq:X2_2d+slab_1}, divided by $z^2$, gives
\begin{eqnarray}
\mathcal{D}_x &=&\frac{\langle \Delta  X^2 \rangle}{z^2} \nonumber \\ &=&\frac{1}{\pi B_0^2} \int_{-\infty}^{\infty}\int_{-\infty}^{\infty}  P_{xx}^{2D}(k_x,k_y) \left[1-\cos\left(k_x X_0 \right)\right]dk_x dk_y. \nonumber \\
\label{eq:curly_Dx2}
\end{eqnarray}
Although Equations~\ref{eq:X2_2d+slab_1} \& \ref{eq:curly_Dx2} do not directly depend on the slab power spectra, $b$ is the root-mean-square of the magnetic fluctuation that also includes the slab component. Therefore, the slab component modifies the separation indirectly through $b$. Equations ~\ref{eq:X2_2d+slab_1} \& \ref{eq:curly_Dx2} are also used to solve for $\left\langle \Delta Y^2 \right \rangle$ and $\mathcal{D}_y$ by replacing $P_{xx}$ with $P_{yy}$.
Because we now assume axisymmetry, we have $\langle \Delta x^2\rangle=\langle \Delta y^2\rangle$ and hence $I_x=I_y$, which we can simply call $I$. 
However, because the initial separation is along $x$ and not along $y$, we obtain different results for $J_x$ and $J_y$, and therefore for $\langle \Delta X^2\rangle$ and $\langle \Delta Y^2\rangle$.

To perform computer simulations of field line trajectories, the forms of $P_{xx}$ and $P_{yy}$ must be specified. We use the axisymmetric turbulence spectra from \citet{Ruffolo2004}:
\begin{eqnarray}
    P_{xx}^{slab}(k_z) &=& P_{yy}^{slab}(k_z) = \frac{\sqrt{2}\Gamma(\frac{5}{6})}{2\Gamma(\frac{1}{3})}\frac{f_{s}\langle b^{2} \rangle l_{z}}{[1+(k_{z} l_{z})^2]^{\nu/2}},
    \label{eq:spec_s} \\
    A(k_\perp) &=& \frac{8}{9}\frac{(1-f_{s})\langle b^{2} \rangle l_{\perp}^{4}}{(1+k_{\perp}^{2} l_\perp^2)^{(\nu+3)/2}},
    \label{eq:spec_a}
\end{eqnarray}
where $l_{z}$ and $l_{\perp}$ are parallel and perpendicular bendover scales, respectively, and 
$\nu$ represents the spectral index of both the 2D omnidirectional magnetic power spectrum at $k_\perp \gg 1/l_\perp$ and the slab magnetic power spectrum at $k_z\gg l_z$.
From these spectra, $\mathcal{D}_x$ and $\mathcal{D}_y$ are 
 \begin{eqnarray}
     \mathcal{D}_{x}&=&\frac{8}{9}\frac{(1-f_{s})b^{2}}{B_{0}^{2}}\left(\frac{2}{\nu^{2}-1}-\frac{(2l_{\perp}/X_{0})^{(1-\nu)/2}K_{\frac{1-\nu}{2}}\left(\frac{X_{0}}{l_{\perp}}\right)}{\Gamma(\frac{3-\nu}{2})}\right), 
     \label{eq:curly_Dx3} \nonumber \\ \\
     \mathcal{D}_{y}&=&\frac{8}{9}\frac{(1-f_{s})b^{2}}{B_{0}^{2}}\left(\frac{2}{\nu^{2}-1}-\frac{(2l_{\perp}/X_{0})^{(1-\nu)/2}K_{\frac{1-\nu}{2}}\left(\frac{X_{0}}{l_{\perp}}\right)}{\Gamma(\frac{3-\nu}{2})} \right.
    \nonumber \\
&& \left. -\frac{2(X_{0}/(2l_{\perp}))^{(1+\nu)/2}K_{\frac{3-\nu}{2}}\left(\frac{X_{0}}{l_{\perp}}\right)}{\Gamma(\frac{3-\nu}{2})}\right),
     \label{eq:curly_Dy3}
 \end{eqnarray}
 where $K_{\alpha}(X_0/l_\perp)$ is the modified Bessel function of the second kind. We then solve Equation~\ref{eq:X2_2d+slab_1} by using the spectra from Equations~\ref{eq:Px2d}-\ref{eq:spec2} and~\ref{eq:spec_s}-\ref{eq:spec_a} and $\mathcal{D}_x$ and $\mathcal{D}_y$ from Equations~\ref{eq:curly_Dx3} and~\ref{eq:curly_Dy3}.
 \begin{figure*}
\includegraphics[width=\textwidth]{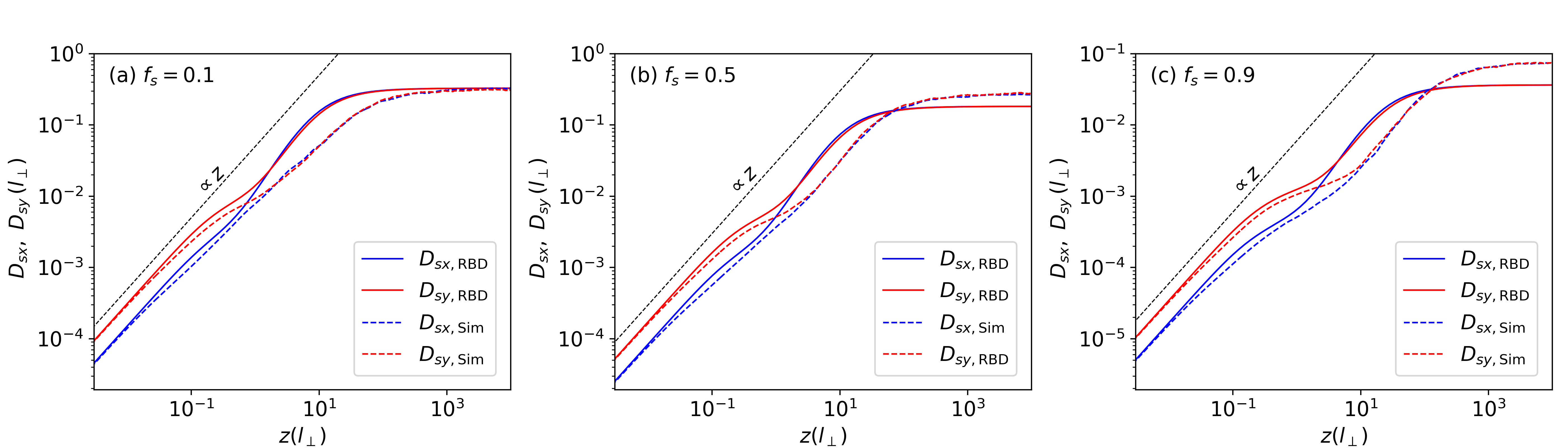}
    \caption{Comparison of diffusion coefficients of field line separation, $D_{sx}$ and $ D_{sy}$, from RBD theory and numerical simulation versus $z$ for $b/B_0=0.5$, $X_{0}/l_{\perp}=0.1$, and $f_s$ equal to (a) 0.1, (b) 0.5, and (c) 0.9. Where $D_{sx}$ or $D_{sy}$ is roughly constant versus $z$, there is a regime of diffusive separation. Note that at large $z$, $D_{sx} = D_{sy} = 2D_{RBD}^{2D}$, i.e., twice the diffusion coefficient of the field line random walk, representing fast diffusive separation of the uncorrelated field lines. For large $f_s$, there is also a regime of slow diffusive separation at moderate $z$ (see text for details). At low $z$, $D_{sx}$ and $D_{sy}$ are very different and both proportional to $z$.}
\label{fig:3cases1}
\end{figure*}
 \begin{figure*}
\includegraphics[width=\textwidth]{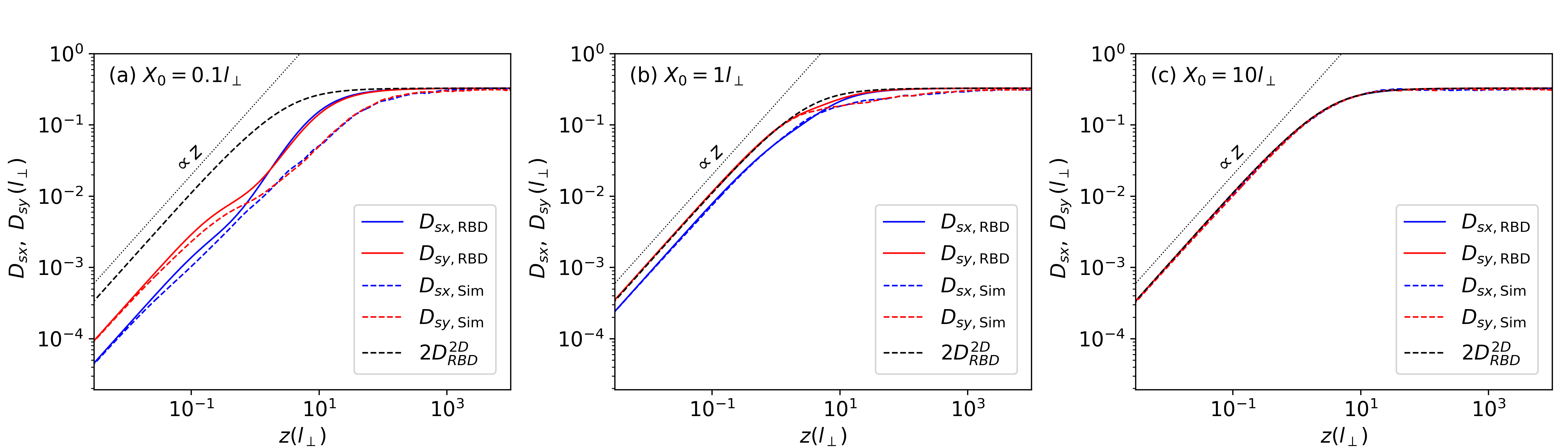}
    \caption{Comparison of $D_{sx}$ and $D_{sy}$ from RBD theory and numerical simulation versus $z$ for $b/B_0=0.5$, $f_s=0.1$, and $X_{0}/l_{\perp}$ equal to (a) $0.1$, (b) $1$, and (c) $10$. As $X_{0}$ increases, the two field lines have fluctuations that are less correlated, so $D_{sx}$ and $D_{sy}$ both tend to $2D^{\rm 2D}_{\rm RBD}$, and there is less difference between them.}
\label{fig:variesX0}
\end{figure*}

From the formulae, $\mathcal{D}_x$ is smaller than $\mathcal{D}_y$. This behavior can be explained if we consider the diffusion of field lines near O-points in the 2D turbulence where the current density $\mathbf{J}\approx J_0 \hat{z}$ is almost constant. The $x$ and $y$ components of the magnetic field near O-points are $b_x=-\mu_{0}J_0y/2$ and $b_y=\mu_{0}J_0 x/2$, respectively. If we consider the change of $b_x$ and $b_y$ from point $(x,y)$ to point $(x+X_0,y)$ near the O-point, there is almost no difference in $b_x$ but $b_y$ changes by $\mu_{0}JX_{0}/2$. Therefore, two field lines separate promptly in the $y$ direction at the beginning while the separation in $x$ starts later and $\mathcal{D}_x<\mathcal{D}_y$.  
For small $X_0$, $\mathcal{D}_{x}$ and $\mathcal{D}_y$ can be written in forms of series: 
\begin{eqnarray}
    \mathcal{D}_{x}&\approx&\frac{8}{9}(1-f_{s})\frac{b^{2}}{B_{0}^{2}}\left[\left(\frac{\Gamma\left(\frac{\nu-1}{2}\right)X_{0}^{2}}{4(\nu-3)\Gamma\left(\frac{\nu+3}{2}\right)} \right.\right. \nonumber \\ & & \left.\left.\mbox -\frac{\Gamma\left(\frac{\nu-1}{2}\right)X_{0}^{4}}{16(\nu-5)(\nu-3)\Gamma\left(\frac{\nu+3}{2}\right)}+\mathcal{O}((X_0)^6)\right)\right. \nonumber \\ & & \mbox -X_{0}^{\nu+1}\left(\frac{\Gamma\left(-\frac{\nu-1}{2}\right)}{2^{\nu}\Gamma\left(\frac{\nu+3}{2}\right)X_{0}^{2}} \right. \nonumber \\ & & \left.\left.\mbox +\frac{\Gamma\left(-\frac{\nu-1}{2}\right)}{2^{\nu+1}(\nu+1)\Gamma\left(\frac{\nu+3}{2}\right)}+ \mathcal{O}((X_{0})^{2})\right)\right], 
    \label{eq:curly_Dx4} \\
    \mathcal{D}_{y}&\approx& \frac{8}{9}(1-f_{s})\frac{b^{2}}{B_{0}^{2}} \left[ \Biggl( \left( \frac{\Gamma(\frac{\nu-3}{2})}{4\Gamma(\frac{\nu+3}{2})} + \frac{\Gamma(\frac{\nu-1}{2})}{4(\nu-3)\Gamma(\frac{\nu+3}{2})} \right) X_{0}^{2} \right.\Biggr. \nonumber \\ 
    & & + \left. \left. 
    \left( \frac{12\Gamma(\frac{\nu-3}{2})-2(\nu+3)\Gamma(\frac{\nu-3}{2})-\Gamma(\frac{\nu-1}{2})}{16(\nu-3)(\nu-5)\Gamma(\frac{\nu+3}{2})} \right) X_{0}^{4} \right. \right. \nonumber \\ 
    & & + \Biggl. \left.\mathcal{O}((X_{0})^{6})  \Biggr) \right. \nonumber \\
    & & + X_{0}^{\nu+1} \Biggl( -\frac{\Gamma(-\frac{\nu-1}{2})}{2^{\nu}\Gamma(\frac{\nu+3}{2})X_{0}^{2}} - \frac{\Gamma(-\frac{\nu-1}{2})}{2^{\nu+1}(\nu+1)\Gamma(\frac{\nu+3}{2})} \Biggr. \nonumber \\ 
    & & - \Biggl. \mathcal{O}((X_{0})^{2}) \Biggr)\nonumber \\
    & & + X_{0}^{\nu+2}\left( \frac{2^{1-\nu}\Gamma(\frac{3-\nu}{2})}{\Gamma(\frac{\nu+3}{2})X_{0}^{3}} + \frac{\Gamma(\frac{3-\nu}{2})}{2^{\nu}(\nu-1)\Gamma(\frac{\nu+3}{2})X_{0}} \right. \nonumber \\ 
    & & + \Biggl.\Biggl. \mathcal{O}((X_{0})^{2}) \Biggr) \Biggr].
    \label{eq:curly_Dy4}
\end{eqnarray}
$\mathcal{D}_{x}$ and $\mathcal{D}_{y}$ are found to be monotonically increasing functions of $X_0$. 
How $\mathcal{D}_x$ and $\mathcal{D}_y$ change, as $X_0$ approaches zero, depends on the spectral index $\nu$.
For $\nu>3$, both $\mathcal{D}_x$ and $\mathcal{D}_y$ are proportional to $X^{2}_0$. When the power spectra have a cutoff wavenumber, the values of $\mathcal{D}_x$ and $\mathcal{D}_y$ will eventually become proportional to $X^{2}_0$ for sufficiently small values of $X_0$. This behavior is illustrated in the field line separation formula provided in \cite{Shalchi2019}, which uses a quasilinear approach to analyze the case of small $X_0$. For $\nu<3$, both $\mathcal{D}_x$ and $\mathcal{D}_y$ are proportional to $X_0^{\nu-1}$. In this work, we use the Kolmogorov index $\nu=5/3$, for which $\mathcal{D}_{x}, \mathcal{D}_{y}\propto X_{0}^{2/3}$ in the limit of small $X_{0}$.

In order to verify our RBD theory, we used the {\tt streamline} code \citep{Dalena2012} to generate representations of 2D+slab turbulence and trace large numbers of field line trajectories. This allowed us to gather statistics on the separation of field lines and compare them with the theory.
We generate field lines inside a rectangular box with $L_x \times L_y \times L_z=204.8l_\perp\times 204.8 l_\perp\times 10^5 l_z$, where $L_x$, $L_y$, and $L_z$ are  the lengths along the $x$, $y$, and $z$ directions.
For our choice of the slab spectrum, the slab correlation length is $l_c=0.747l_z$, which we set equal to $l_\perp$; thus the value of $l_z$ is $1.339l_\perp$. 
The resolution of 
$N_x\times N_y\times N_z=$ $2^{12}\times2^{12}\times2^{23}$=4,096 $\times$ 4,096 $\times$ 8,388,608 is used unless otherwise stated.
This results in grid spacings of $\Delta x=\Delta y=0.05l_\perp$ and $\Delta z=0.01192l_z$.
The 2D+slab turbulent magnetic fields are generated using inverse fast Fourier transforms from the spectra given in Equations~\ref{eq:spec1}-\ref{eq:spec2} and~\ref{eq:spec_s}-\ref{eq:spec_a}, and the phase of each Fourier mode is random. The resulting fields at grid points are normalized to obtain the desired rms values. Five thousand pairs of field lines were generated for each simulation setup.
We solved the field line equations by using the Cash–Karp method \citep{Cash&Karp1990,Press1992} and a fifth-order Runge-Kutta method with adaptive step size.
For comparison purposes, we also calculate $\langle \Delta X^2\rangle$ and $\langle \Delta Y^2\rangle$ 
using the DD theory of \cite{Ruffolo2004} as modified according to Footnote 4 of \cite{Ruffolo2015} to set $z''>z'$ in derivation. 

Figure~\ref{fig:3cases1} shows $D_{sx}=\langle \Delta X^2 \rangle /(2z)$ and $D_{sy}=\langle \Delta Y^2 \rangle /(2z)$ from RBD and the simulations.
We use $b=0.5B_0$, $X_0=0.1l_{\perp}$, and $f_s=b^2_{slab}/b^2=0.1$, $0.5$, and $0.9$. We note that since discrete Fourier modes are used in the simulation, we also modify Equation~\ref{eq:X2_2d+slab_1} to sum over the same discrete values of $(k_{x},k_{y})$.
\begin{figure*}
\includegraphics[width=\textwidth]{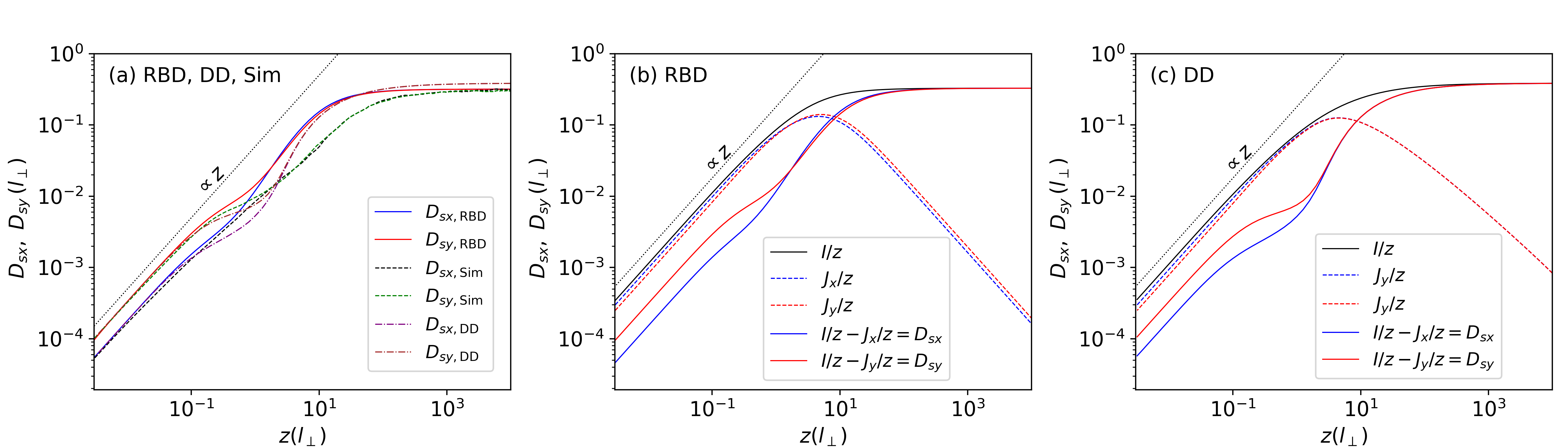}
    \caption{(a): Comparison of $D_{sx}$ and $D_{sy}$ from RBD theory, DD theory, and numerical simulation versus $z$. 
    (b) and (c): $D_{sx}=I/z-J_x/z$ and $D_{sy}=I/z-J_y/z$ and their component terms as functions of $z$, for (b) RBD and (c) DD theories. 
    All panels are for $b/B_0=0.5$, $f_s=0.1$,  and $X_{0} = 0.1l_{z}$, and the black dotted line is proportional to $z$.}
\label{fig:3cases}
\end{figure*}

At $z<10^{-2}l_\perp$, $D_{sx}$ and $D_{sy}$ from both RBD theory and simulations match very well. Specifically, $D_{sx}$ and $D_{sy}$ are  $\mathcal{D}_xz/2$ and $\mathcal{D}_yz/2$ respectively, in line with the predictions made in Equations~\ref{eq:actual_X2} and~\ref{eq:actual_Y2}. Physically, this good agreement occurs because the field line displacement is proportional to $z$ at very small $z$. At $10^{-2}l_\perp<z<l_\perp$, there is some discrepancy between RBD theory and simulation results. We find that these differences can be substantially reduced by improving the resolution in the simulation; results for other $z$ ranges are not sensitive to the resolution.   

For all $f_s$ and at $l_\perp <z < 10^2 l_\perp$, RBD theory and simulations give different $D_{sx}$ and $D_{sy}$ values.  
At these scales of $z$, the RBD assumption becomes less accurate for large $\Delta z'$ where $\sigma^2_X$ and $\sigma^2_Y$ should instead be proportional to $\Delta z'$. This causes $J_{x}$ to drop at lower $z$ and $D_{sx}$ and $D_{sy}$ as predicted by RBD are larger than the values from the simulations. From Figure~\ref{fig:3cases1}, RBD theory gives a very good approximation as $f_s$ becomes small. With increasing $f_{s}$, there is a wider plateau region near $z\sim l_{\perp}$, which will be discussed shortly. Figure~\ref{fig:variesX0} compares $D_{sx}$ and $D_{sy}$ from RBD theory and the simulations for $f_s=0.1$ and $X_{0}/l_{\perp}=0.1$, $1$, and $10$. 
As $X_0$ increases, RBD gives $D_{sx}$ and $D_{sy}$ that match very well with the result from the simulations at all $z$. 

For very large $X_0$, we can neglect the term $J_{x}$ because $\cos(k_x X_0)$ rapidly oscillates and $J_{x}$ becomes very weak.
As $X_0$ goes to infinity, we obtain
\begin{equation}
\left\langle \Delta X^2 \right \rangle =\left\langle \Delta Y^2 \right \rangle =2\langle \Delta x^2\rangle^{\rm 2D}=2\langle \Delta y^2\rangle^{2D}=4D^{\rm 2D}_{\rm RBD}z,
\end{equation}
where
\begin{eqnarray}
D^{\rm 2D}_{\rm RBD}
&=& \frac{1}{\sqrt{\pi} b B_0}\int_{-\infty}^{\infty}\int_{-\infty}^{\infty}  \frac{P_{xx}^{\rm 2D}(k_x,k_y)}{k_\perp} dk_x dk_y. 
\label{eq:dX2_2d_zbig}
\end{eqnarray}

Indeed, for axisymmetric turbulence, $D_{sx}=D_{sy}=2D^{2D}_{x}=2D^{\rm 2D}_{\rm RBD}$. 
Replacing $P^{\rm 2D}_{xx}$ by $P^{2D}_{yy}$ in Equation~\ref{eq:dX2_2d_zbig} gives the same $D^{\rm 2D}_{\rm RBD}$.
In this limit of large $X_{0}$, there is no 2D correlation between $\mathbf{b}$ at the two field lines, which therefore undergo two independent 2D random walks. Our theory and simulation results do exhibit $D_{sx}=D_{sy}=2D^{\rm 2D}_{\rm RBD}$ when $X_{0}=10l_{\perp}$, as shown in Figure~\ref{fig:variesX0}(c). Conversely, for any $X_{0}$, the two field lines eventually become far apart, at very large $z$, so the 2D random walks become uncorrelated and $D_{sx}=D_{sy}=2D^{\rm 2D}_{\rm RBD}$, as is also seen in Figure~\ref{fig:variesX0}. Then at large $z$, $D_{sx}=D_{sy}$ becomes constant, so the field line separation is diffusive, what \cite{Ruffolo2004} termed fast diffusive separation. 

Figure~\ref{fig:3cases}(a) shows $D_{sx}$ \& $D_{sy}$ from RBD, DD and simulations, as well as the behavior of terms $I_{x}$ and $J_{x}$.
The resolution used here is $N_x\times N_y\times N_z=$ 8,192 $\times$ 8,192 $\times$ 8,388,608, with the same simulation box size.
At $z<10^{-2}l_\perp$, all theories and simulations give very similar results. At $10^{-2}l_\perp< z <l_\perp$, neither theory can predict $D_{sx}$ or $D_{sy}$ well. For this specific setup with $f_s=0.1$, DD gives $D_{sx}$ and $D_{sy}$ that are a little too high compared to those from RBD and simulations. 

To better understand the behavior of $D_{sx}$ and $D_{sy}$ from RBD, let us consider $d\langle \Delta X^2 \rangle /dz$ and $d\langle \Delta Y^2\rangle/dz$ because they are simpler and $D_{sx}\sim d\langle \Delta X^2\rangle/dz$ or $D_{sy}\sim d\langle \Delta Y^2\rangle/dz$ if we consider only the order of magnitude.
The derivative of Equation~\ref{eq:X2_2d+slab_1} with respect to $z$ is
\begin{equation}
\frac{d\langle \Delta X^2\rangle}{dz} = 2\dot{I}-2\dot{J}_x,
\end{equation}
where
\begin{eqnarray}
    \dot{I} &=& \frac{1}{\sqrt{\pi}bB_{0}} \int  \frac{P_{xx}^{2D}({\bf k}_\perp)}{k_\perp}{\rm erf}\left( \frac{bk_\perp}{2B_{0}}z \right) d{\bf k}_\perp \\
    \dot{J}_x &=& \frac{1}{\sqrt{\pi}B_{0}^2} \int \frac{P_{xx}^{2D}({\bf k}_\perp) \cos(k_{x}X_{0})}{\sqrt{(bk_{\perp}/B_{0})^{2}+2(\mathcal{D}_{x}k_{x}^{2}+\mathcal{D}_{y}k_{y}^{2})}} \nonumber \\
    & & \times \exp\left(-\frac{(bk_{\perp}/B_{0})^{2}(\mathcal{D}_{x}k_{x}^{2}+\mathcal{D}_{y}k_{y}^{2})z^{2}}{2(bk_{\perp}/B_{0})^{2}+4(\mathcal{D}_{x}k_{x}^{2}+\mathcal{D}_{y}k_{y}^{2})}\right)  \nonumber \\
    & & \times \left[ {\rm erf}\left( \frac{(\mathcal{D}_{x}k_{x}^{2}+\mathcal{D}_{y}k_{y}^{2})z}{\sqrt{(bk_{\perp}/B_{0})^{2}+2(\mathcal{D}_{x}k_{x}^{2}+\mathcal{D}_{y}k_{y}^{2})}} \right) \right. \nonumber \\
    & & + \left. {\rm erf}\left( \frac{(1/2)(bk_{\perp}/B_{0})^{2} z}{\sqrt{(bk_{\perp}/B_{0})^{2}+2(\mathcal{D}_{x}k_{x}^{2}+\mathcal{D}_{y}k_{y}^{2})}} \right) \right] d{\bf k}_\perp. \nonumber \\
\end{eqnarray}

The field line separation is controlled by the the power spectra and how $\dot{I}$ and $\dot{J}_x$ change as functions of $z$. First, we consider the case of small $X_0$.  Here $\dot{I}$ and $\dot{J}_x$ are very similar at small $z$. Both terms are proportional to $z$. If their behaviors become slightly different at a given value of $z$, then $d\langle \Delta X^2\rangle/dz$ can change significantly at that $z$.
At $z\sim z_1 \approx 2B_0 l_\perp/b$ (where $\langle\Delta x^2\rangle^{2D}\approx l_\perp^2$) or even an order less, the term $\dot{I}$ begins to deviate from being proportional to $z$. This is because the error function in $\dot{I}$ starts to saturate and tend toward unity. However $\dot{J}_x$ is still increasing with $z$ at $z\sim z_1$ because there are two error functions in $\dot{J}_x$ and only of them begins to saturate at $z\sim z_1 \approx 2B_0 l_\perp/b$, while the other begins to saturate at $z\sim z_2 \approx (b^2/B^2_0) l_\perp/\mathcal{D}_{y}$, noting that $\mathcal{D}_y>\mathcal{D}_x$ so $\mathcal{D}_y$ is dominant. For small $X_0$, $z_2$ is greater than $z_1$. Therefore at $z_1\lesssim z\lesssim z_2$, 
$d\langle \Delta X^2\rangle/dz$ and $d\langle \Delta Y^2\rangle/dz$ can increase more slowly as $z$ increases. This gives rise to a plateau where $D_{sx}$ and $D_{sy}$ are independent of $z$, implying diffusion behavior that was termed slow diffusive separation by \cite{Ruffolo2004}.
Note that a clear regime of slow diffusive separation, as shown in Figure 8 of \cite{Ruffolo2004}, appears for sufficiently small $X_0$ (see Figures~\ref{fig:3cases1} and \ref{fig:variesX0}(a)).  For moderately small $X_0$, as shown in Figure~\ref{fig:3cases}, there is a hint of slow diffusive separation over $z/l_\perp\sim0.2$ to 1.
For RBD, if $f_s$ is closer to unity and very small $X_0$ is used, $z_2$ should be much greater than $z_1$ and both $d\langle \Delta X^2\rangle/dz$ and $d\langle \Delta Y^2\rangle/dz$ can be close to constant over a wider range $z_1<z<z_2$, the slow diffusive regime seen in Figure~\ref{fig:3cases1}(c). 

At $z\sim z_2$ or an order less, $\dot{J}_x$ drops as all error functions are saturating toward unity and the exponential term begins to drop so that $d\langle \Delta X^2\rangle/dz$ increases more rapidly than being proportional to $z$. In  this range of $z$, the field line separation is superdiffusive. 

At large $X_0$, the term $\dot{J}_x$ is tiny because the integrand rapidly oscillates, and the integral tends toward zero. Therefore both $\langle \Delta X^2\rangle$ and $\langle \Delta Y^2\rangle$ are very similar and close to $2 I$ as shown in Figure~\ref{fig:variesX0}(c), corresponding to fast diffusive separation.

Figures~\ref{fig:3cases}(b) \&~\ref{fig:3cases}(c) show $I/z$ and $J_{x}/z$ for $\langle \Delta X^2\rangle$ and $\langle \Delta Y^2\rangle$ from RBD and DD theory. 
Both theories provide similar behavior for $I$ and $J_{x}$. A slight change of behavior of $I$, possibly starting when $z$ is one order of magnitude less than $z_1$, causes the slow diffusion. Once $z$ increases and become close to $z_2$, the diffusion becomes fast as $J_{x}$ drops toward zero.

\begin{figure}
\hspace*{-1.5cm}\includegraphics[width=4.5in]{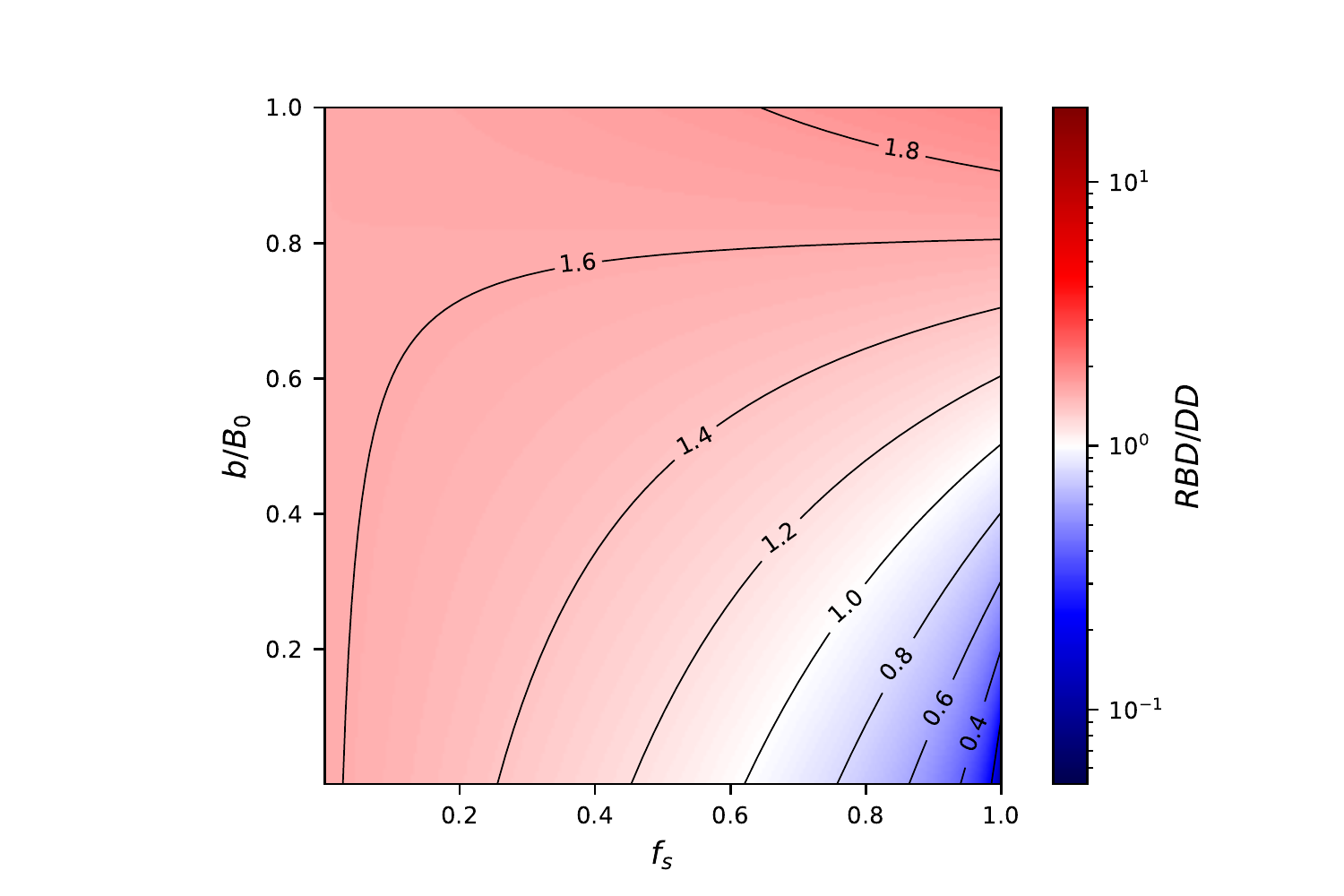}
    \caption{A 2D color plot of $D_{sx,\rm{RBD}}/D_{sx,\rm{DD}}$ at large $z$ for varying $f_{s}$ and $b/B_{0}$.}
\label{fig:bB0vsfs}
\end{figure}

We can compare RBD and DD at $z\gg l_\perp$.
Integrating equations \ref{eq:dX2_2d_zbig} over $k_{x}$ and $k_{y}$ for $\nu = 5/3$ gives
\begin{eqnarray}
    D_{sx,{\rm RBD}} &=& D_{sy,{\rm RBD}} = 2D^{\rm 2D}_{\rm RBD} = \frac{\pi \Gamma(\frac{5}{6})}{\Gamma(\frac{1}{3})}(1-f_{s}) \frac{b}{B_{0}}l_{\perp} \nonumber\\
    &=& \sqrt{\pi}(1-f_{s})\frac{b}{B_{0}}\lambda_{c2},
\end{eqnarray}
where $\lambda_{c2}$ is the total correlation length of the 2D turbulence \citep{Matthaeus2007, Ghilea2011}.
The expressions of $D_{sx}$ and $D_{sy}$ at large $z$ from the DD theory \citep{Ruffolo2004} are
\begin{equation}
    D_{sx,{\rm DD}} = D_{sy,{\rm DD}} = 2\frac{(D_\perp^{2D})^2}{D_{\perp}} = \frac{\tilde{\lambda}^{2}}{D_{\perp}}(1-f_{s}) \frac{\langle b^{2} \rangle}{B_{0}^{2}},
\end{equation}
where $\tilde{\lambda}$ is the ultrascale, $D_\perp=(D_\perp^{slab}/2) + \sqrt{(D_\perp^{slab}/2)^2 + (D_\perp^{2D})^2}$, $D_{\perp}^{slab}=(1/2)f_{s}(\langle b^2 \rangle/B_{0}^2)l_{\perp}$, and $D_{\perp}^{2D}=(\tilde{\lambda}/\sqrt{2})\sqrt{1-f_{s}}(b/B_{0})$. Figure \ref{fig:bB0vsfs} shows the ratio of $D_{sx,{\rm RBD}}$ to $D_{sx,{\rm DD}}$ as a function of $b$ and $f_{s}$ as $z$ goes to infinity for our spectra. The ratio can be less or greater than unity. The ratio is between $0.053$ and $1.97$. At $b/B_{0}\gtrsim 0.5$, the ratio is always greater than unity for any value of $f_{s}$. For $b/B_{0}\gtrsim 0.5$, the ratio can be less than unity at $f_{s}$ close to 1. For lower $b/B_{0}$, the lower limit of $f_{s}$ that gives the ratio less than unity becomes lower. 
The white region gives the ratio equal to unity. In this region, both DD and RBD predict the same field line separation at large $z$. In the limit $f_{s}\rightarrow 0$, the ratio becomes $1.62$ for any $b/B_{0}$.

\begin{figure}
\hspace*{-0.5cm}\includegraphics[width=4in]{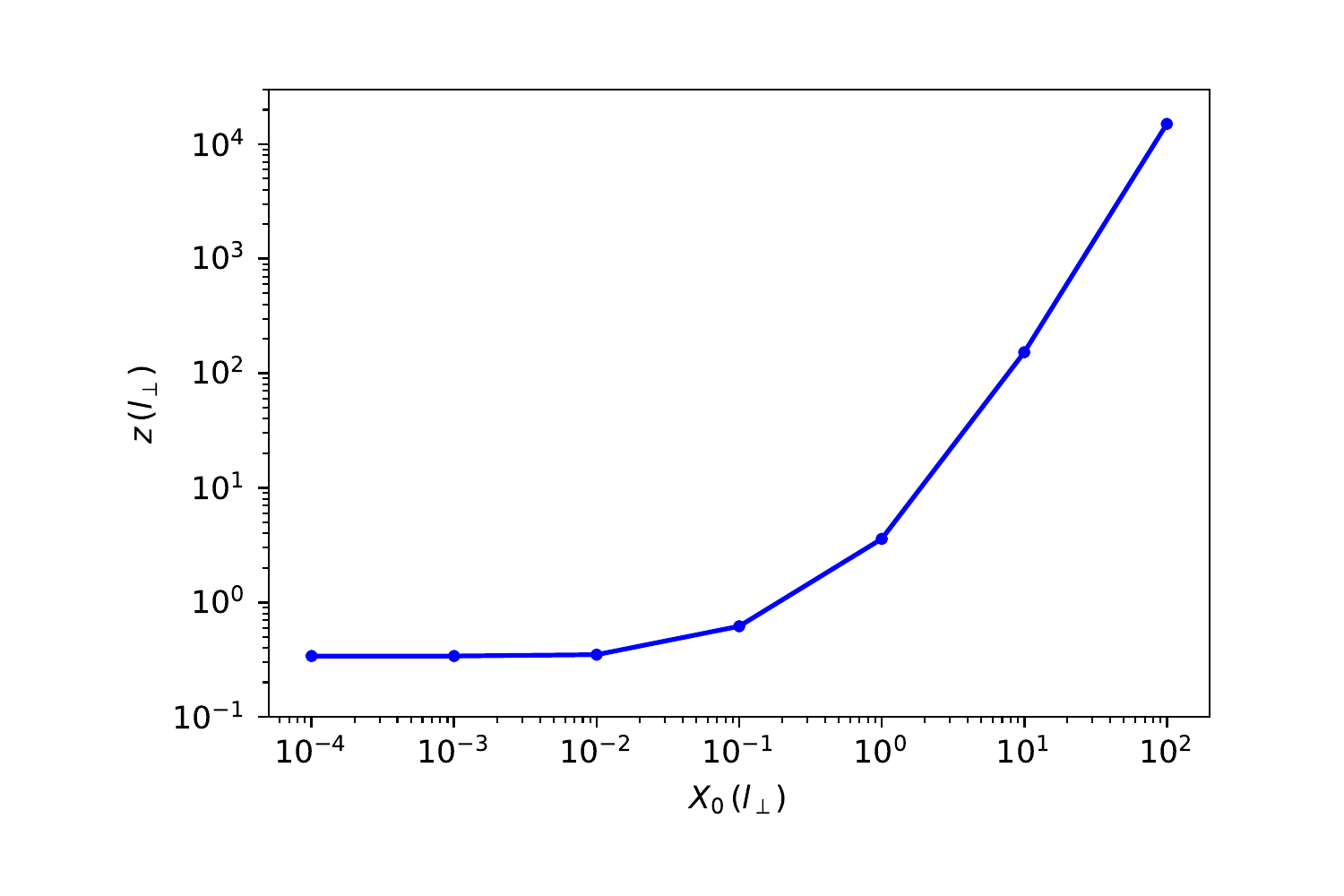}
    \caption{The value of $z_0$ where $\langle\Delta X^{2}\rangle=X_{0}^{2}$, as a function of $X_{0}$. This represents the distance along the mean field after which spatial structure becomes dispersed. Here $b/B_{0}=0.5$ and $f_{s}=0.1$.}
\label{fig:X0vsz}
\end{figure}

All plots of RBD and DD results are for discrete spectra because we want to compare them with the simulations. We investigate the effect of discrete 2D spectra in the simulations by increasing $N_{x}$ and $N_{y}$ from 4096 to 8192. We found that with better resolution, $D_{sx}$ and $D_{sy}$ from simulations move closer to the RBD and DD results at small $z$, but at $z\gtrsim z_1$ the difference remains essentially unchanged. We conclude that if we continue to improve the resolution, simulation, RBD, and DD should give the same results at small $z$.

\section{Discussion and Conclusions}

We have derived the formulae to estimate how two field lines separate on average as a function of distance $z$ parallel to the mean magnetic field using the RBD assumption. We adopt Corrsin's hypothesis to account for the probability distribution of field line displacements and separations. The distributions are assumed to be Gaussian and RBD is used to specify $\sigma_x$, $\sigma_y$, $\sigma_X$, and $\sigma_Y$ as a linear functions of $z$.
We compare the statistics from RBD, DD, and simulations for 2D+slab power spectra. Our results show how the field line separation is controlled by the slab fraction $f_{s}$, the initial separation $X_0$, and magnetic turbulence amplitude $b$. The terms $I_{x}=\langle \Delta x^{2}\rangle$ and $J_{x}$ control the shapes of $D_{sx}$. 
The field line separation from RBD and simulations share very similar behaviors. They become different at $z\sim l_\perp$ where the RBD should give an inaccurate $\mathcal{D}_x$ and $\mathcal{D}_y$, a regime where DD also cannot provide a good approximation.
The field line separation based on RBD provides a more accurate description for low $f_s$, as typically found in near-Earth interplanetary space. 

The field line separation can exhibit four regimes, which are free streaming, slow diffusive separation, superdiffusive separation, and fast diffusive separation. 
At low $z$, for free streaming, both RBD and DD theories give simple formulae for $\langle \Delta X^2 \rangle$ and $\langle \Delta Y^2 \rangle$ which can be derived without using any assumption about the spread of field lines, matching the simulation results. This regime is superdiffusive in the sense that $D_{sx}$ and $D_{sy}$ are increasing with $z$. But when $D_{sx}$ or $D_{sy}$ is nearly constant in $z$, the field line separation is diffusive.
For example, for $z_{1} \lesssim z \lesssim z_{2}$ and small $X_{0}$, we obtain slow diffusive separation, i.e., a plateau at a lower diffusion coefficient than for the fast diffusive separation at large $z$ (see Figures~\ref{fig:3cases1} and~\ref{fig:variesX0}(a)). It only becomes obvious for $f_s$ close to unity and small $X_0$ (Figure~\ref{fig:3cases1}(c)). In such cases, at $z\gtrsim z_{2}$, there is another regime of superdiffusive separation in which $D_{sx}$ and $D_{sy}$ rise to the level of fast diffusive separation, at $\langle \Delta x^{2} \rangle^{\rm 2D} \gg l_{\perp}^{2}$ with $D_{sx}=D_{sy}=2D_{\rm RBD}^{2D}$ that is nearly constant at its asymptotic value.
Physically this represents the separation between two field lines with independent 2D random walks, each with diffusion coefficient $D_{\rm RBD}^{2D}$.

At very large $z$, results for $D_{x}$ based on RBD, DD, and ODE theories and simulations are known to differ for 2D+slab \citep{Ghilea2011,Snodin2013a,Snodin2016} and for isotropic turbulence \citep{Sonsrettee2015,Sonsrettee2016}. For 2D+slab, except for the cases of a very weak slab contribution, RBD, DD, or ODE could all be used to describe
the direct simulation results. 
For strong and weak slab contributions, RBD and ODE theory are effective at matching the simulation results. Even if the ODE theory provides a better match than RBD in some cases, the explicit formula for RBD is more convenient. For isotropic turbulence, RBD, DD, and ODE can qualitatively describe simulation results at any $b/B_0$. When the fluctuations are fully transverse polarized in real space ($b_z =0$) but still isotropic in $k$-space, at any $b/B_0$ range RBD provides the best match to $D_{x}$ in the simulation results \citep{Sonsrettee2016}.

Assuming that plasma tied to a bunch of field lines carries some information about its location of origin, such as the presence of energetic particles or plasma composition, this information must intermix or become blurred as $z$ increases due to the diffusive or superdiffusive field line separation. To investigate how fast the information at the scale $X_0$ become dispersed, in Figure~\ref{fig:X0vsz} we plot the distance $z_0$ along the mean field where $\langle \Delta X^2\rangle$ equals $X^2_0$ for $b/B_{0}=0.5$ and $f_{s}=0.1$. As $X_0$ becomes larger, $z_0$ also increases, but as $X_{0}\rightarrow 0$, we find $z_{0}\rightarrow 0.35l_\perp$. We interpret this as follows.
Any information with spatial scales less than $0.01l_\perp$ can persist (at that scale) up to the distance $z_0=0.35l_\perp$. Indeed, this value relates to the scale of exponential separation between chaotic field lines, which is manifest as the free-streaming regime at low $z$ \citep{Ruffolo2004}. However, information at spatial scales greater than $0.01l_\perp$ can persist to a greater distance. This phenomenon was demonstrated in simulations concerning dropouts of solar energetic particles \citep{Ruffolo2003,Chuychai2007,Tooprakai2016}. Structure at spatial scales smaller than typical 2D magnetic islands (of size $\sim l_{\perp}$) was dispersed at a shorter distance $z$ than larger structures. 

The numerical calculation of the explicit formulae derived here using RBD is very straightforward compared with the implicit formulae derived by \cite{Ruffolo2004} using DD. The equations for solving $D_{sx}$ and $D_{sy}$ in the DD model are implicit and display significant oscillation. To obtain values for $D_{sx}$ and $D_{sy}$, we need to determine the roots of these equations, which involves guessing initial values for $D_{sx}$ and $D_{sy}$. This guessing process can be challenging. One effective approach is to use the values of $D_{sx}$ and $D_{sy}$ obtained from the RBD model as initial guesses when solving for $D_{sx}$ and $D_{sy}$ in the DD model. The formulae derived here can also be used with a general form of transverse magnetic power spectra. The field line separation using RBD theory should be a useful tool in various fields of study involving magnetic turbulence.

\section*{Acknowledgements}

This work was supported by the Office of the Permanent Secretary, Ministry of Higher Education, Science, Research and Innovation  (OPS MHESI) and Thailand Science Research and Innovation (TSRI) (Grant RGNS 63-045). We also acknowledge support from Thailand Science Research and Innovation (RTA6280002) and from Thailand's National Science and Technology Development Agency (NSTDA) and National Research Council of Thailand (NRCT): High-Potential Research Team Grant Program (N42A650868).

\section*{Data Availability}
Our research solely relies on analytical and numerical calculations and does not involve any observational data.  The simulation data and codes used in our study will be provided upon request.
 



\bibliographystyle{mnras}







\bsp	
\label{lastpage}
\end{document}